\documentclass[prd,preprint,nofootinbib]{revtex4}
\usepackage{graphicx}
\usepackage{amssymb}
\usepackage{bm}

\newcommand{\bear}{\begin{array}}  \newcommand{\eear}{\end{array}}
\newcommand{\bea}{\begin{eqnarray}}  \newcommand{\eea}{\end{eqnarray}}
\newcommand{\beq}{\begin{equation}}  \newcommand{\eeq}{\end{equation}}
\newcommand{\bef}{\begin{figure}}  \newcommand{\eef}{\end{figure}}
\newcommand{\bec}{\begin{center}}  \newcommand{\eec}{\end{center}}
\newcommand{\non}{\nonumber}

\newcommand{\la}{\left\langle} \newcommand{\ra}{\right\rangle}

\newcommand{\be}{\begin{eqnarray}}
\newcommand{\ee}{\end{eqnarray}}

\def\FIG#1{Fig.~\ref{#1}}
\def\EQ#1{Eq.~(\ref{#1})}

\begin{document}

\title{The oscillation effects on thermalization of the neutrinos
 in the universe with low reheating temperature}

\author{Kazuhide Ichikawa, Masahiro Kawasaki and Fuminobu Takahashi}
\affiliation{Institute for Cosmic Ray Research,
University of Tokyo, Kashiwa 277 8582, Japan}

\date{\today}

\vskip10mm
\begin{abstract}
We study how the oscillations of the neutrinos affect their
thermalization process during the reheating period with temperature
$O(1)$ MeV in the early universe. We follow the evolution of the
neutrino density matrices and investigate how the predictions of big
bang nucleosynthesis vary with the reheating temperature. For the
reheating temperature of several MeV, we find that including the
oscillations makes different predictions, especially for $^4$He
abundance. Also, the effects on the lower bound of the reheating
temperature from cosmological observations are discussed.
\end{abstract}
\maketitle

\section{Introduction} \label{sec:introduction}
The standard big bang model assumes that the universe was once
dominated by thermal radiation composed of photons, electrons,
neutrinos, and their anti-particles. It is one of the main issues in
theories beyond the standard cosmology where these particles came
from, or equivalently, what reheated the universe.  The reheating
temperature, at which the universe becomes radiation-dominated, is
therefore a very important parameter that discriminates among many
scenarios on the thermal history of the universe. In the following we
define the reheating temperature as that of the latest reheating
process, if the universe experienced several reheating stages.

Recent observations of the cosmic microwave background radiation (CMB)
\cite{Spergel:2003cb} has strongly suggested that the universe
underwent inflation at an early stage.  After inflation ended, the
universe was dominated by the oscillation energy of the inflaton until
it decayed and reheated the universe. The upper limit on the reheating
temperature was obtained \cite{Kawasaki:2004yh} by constraining the
relic abundance of the gravitinos, the superpartner of the graviton,
which are inevitably present in the supersymmetric (SUSY) framework.
Here we are interested in the relatively low reheating temperature,
especially in the MeV range, and would like to put a lower limit on
the reheating temperature.

The MeV-scale reheating is actually ubiquitous in theories beyond the
standard cosmology.  In the framework of the SUSY and superstring
theories, there are many particles with very long lifetime, e.g., the
moduli and the gravitinos mentioned above, since their interaction is
so weak, typically suppressed by the Planck scale.  These long-lived
massive particles might have dominated over the radiation from the
inflaton decay.  If the masses of these particles are heavy enough,
they decay and reheat the universe again just before the big bang
nucleosynthesis (BBN) starts. Otherwise they often cause cosmological
disaster known as ``cosmological moduli problem"
\cite{Coughlan:1983ci,Banks:1993en,deCarlos:1993jw} and 
``gravitino problem" \cite{Kawasaki:2000qr, Jedamzik:1999di,
Cyburt:2002uv, Kawasaki:2004yh}.  The simplest solution of these
problems is to dilute the unwanted relics by producing large entropy
at a later time \cite{Lyth:1995hj,Kawasaki:2004rx}. In either case,
the reheating temperature is very low and typically around MeV.

Another example that prefers the low reheating temperature is the
curvaton scenario~\cite{curvaton} in which the curvaton field
dominates the universe and its isocurvature fluctuation is transformed
into an adiabatic one.  Furthermore, in the Affleck-Dine mechanism
\cite{Affleck:1984fy} responsible for the origin of the baryon
asymmetry, it is known that non-topological solitons such as $Q$-balls
\cite{qball}  are
generally created. Since the decay process of the $Q$-balls is
geometrically suppressed, they might dominate the universe, and such
possibility has been extensively studied in many different scenarios
\cite{Kawasaki:2002hq,Kasuya:2003va,Hamaguchi:2003dc,Ichikawa:2004pb}.

What if the reheating temperature is several MeV? In contrast to
electrons that are always (at least until the temperature drops below
a few eV) in thermal contact with photons via electromagnetic forces,
neutrinos interact with electrons and themselves only through the weak
interaction. The decoupling temperature of the neutrinos should be
around $ 3$ MeV for the electron neutrinos and $5$ MeV for the muon
and tau neutrinos, respectively
\cite{Hannestad:1995rs,Dolgov:1997mb,Dolgov:1998sf}.  The difference
comes from the fact that the electron neutrinos have additional
charged current interaction with electrons. Therefore the neutrinos
might not be fully thermalized if the reheating temperature is in the
MeV range. If this is the case, the expansion rate of the universe
becomes smaller, which affects the light element abundances and the
CMB angular power spectrum
\cite{Kawasaki:1999na,Kawasaki:2000en,Hannestad:2004px,Gelmini:2004ah}.  In
particular it has been widely believed or taken for granted that the
predicted abundance of $^4$He decreases as the reheating temperature
drops below a few MeV. This is because the smaller expansion rate
delays the decoupling of the neutron-proton transformation, decreasing
the neutron-to-proton ratio at the beginning of the BBN. Since almost
all the neutrons are absorbed in the $^4$He nuclei, such a naive
reasoning can explain the dependence of the $^4$He abundance on the
reheating temperature.  In this paper, however, we will see that this
widespread picture is drastically changed if we take account of the
neutrino oscillations.
 
Recent neutrino oscillation experiments~\cite{atom_nu,solar_nu} have
determined the mass differences and mixing angles with high precision
and established that mixing angles are large.  The crucial point is
that a flavor eigenstate transforms itself into another one. Therefore
we must take special care to calculate neutrino distribution functions
and the resultant effective number of neutrinos.  As pointed out in
Refs.~\cite{Dolgov:1980cq, Rudzsky90, McKellar:1992ja,Sigl:1992fn}, it is useful to follow the evolution of the
neutrino density matrices when flavor mixings are present.  Here we
will solve momentum dependent Boltzmann equations for the neutrino
density matrices. 
We will see that the predicted
abundance of $^4$He is drastically changed, while the effective number
of neutrinos does not change much. To put it simply, the reason for
this is that the number density of the electron neutrinos is decreased
due to flavor mixings, which makes the freeze-out temperature of the
neutrons higher; this effect cancels and even overcomes that of the
decrease in the expansion rate. Thus MeV-scale reheating scenario is
one of the examples in which the neutrino oscillations play a
essential role.

The outline of this paper is the following.  In the next section we
formulate neutrino thermalization including flavor mixings, and derive
a evolution equation of the neutrino density matrix.  In the
section~\ref{sec:implications} we will show how the predicted
abundances of the light elements are modified when the reheating
temperature is in the MeV range, and discuss their
implications. Finally we present our conclusion in the
section~\ref{sec:conclusion}.

\section{Neutrino thermalization} \label{sec:thermalization}
In this section, we illustrate the formulation needed to follow the
neutrino thermalization process. The case without the neutrino
oscillations is studied by Refs.~\cite{Kawasaki:1999na,
Kawasaki:2000en, Hannestad:2004px}. Although subjects of study are
different from this paper, issues of the neutrino spectrum evolution
using momentum dependent Boltzmann equations in the early universe are
treated in Refs.~\cite{Hannestad:1995rs, Dolgov:1997mb, Dolgov:1998sf,
Esposito:2000hi, Mangano:2001iu, Dolgov:2002ab,Ichikawa:2003ai}. Our
formulation almost goes in parallel with the no-mixing case and we use
some of the useful techniques discussed in those papers. However,
there is a very important exception that neutrino distribution
functions have to be extended to neutrino density matrices
\cite{Sigl:1992fn} in order to include oscillations.

First of all, let us explain our assumptions on the reheating process
and the neutrino oscillations. We refer to the massive particles which
reheat the universe, or inflaton, as $\phi$~\footnote{
Hereafter we call $\phi$ as inflaton even if it is not responsible for inflation.
}. We assume $\phi$ only decays into photons (they in turn produce
electrons and positrons and they all thermalize very quickly by the
electromagnetic interaction). In other words, the branching ratios to
neutrinos or hadrons are assumed to be negligible and neutrinos are
produced exclusively via the electron-positron annihilation. Then
$\phi$ is characterized simply by its decay rate $\Gamma$. We
parametrize it by the reheating temperature $T_R$ which is defined as
\begin{eqnarray}
\label{eq:def_trh}
\Gamma = 3 H (T_R),
\end{eqnarray}
where $H$ is the expansion rate of the universe. Here, we use the
Friedmann equation $H^2 = \rho_{\rm tot}/3 M_{\rm Pl}$ where the
reduced Planck mass $M_{\rm Pl} = 2.435 \times 10^{18}$ GeV. The total
energy density $\rho_{\rm tot}$ which consists of the radiation
including photons, electrons, and three species of neutrinos, is
expressed as $\rho_{\rm tot} = (g_{\ast} \pi^2/30) T_R^4$ where the
relativistic degree of freedom $g_{\ast} = 43/4$.  This leads to
\begin{eqnarray}
\label{eq:def_trh2}
\Gamma = 3.26 \frac{T_R^2}{M_{\rm Pl}}
    = 2.03 \left(\frac{T_R}{{\rm MeV}} \right)^2 ~ {\rm sec}^{-1}.
\end{eqnarray}
It should be noted that, even if the neutrinos are not fully thermalized, we stick to
\EQ{eq:def_trh2} as the definition of $T_R$ to avoid unnecessary confusion.

We consider three active flavors of neutrinos: $\nu_e$, $\nu_\mu$ and
$\nu_\tau$. When the oscillations are neglected as in
Refs.~\cite{Kawasaki:1999na, Kawasaki:2000en, Hannestad:2004px}, there
are only two sets of variables required to describe the neutrino
evolution. They are the distribution functions for $\nu_e$ and
$\nu_\mu$ which have to be distinguished since they interact
differently with electrons; $\nu_e$ interacts via both neutral and
charged currents while $\nu_\mu$ and $\nu_\tau$ has only the former
interaction. Since $\nu_\mu$ and $\nu_\tau$ interact with electrons
identically, we do not need to solve for the distribution function of
$\nu_\tau$. It is same as $\nu_\mu$'s. On the contrary, when we
include the oscillations among them, $\nu_\mu$ and $\nu_\tau$ also
have to be distinguished because their oscillations between $\nu_e$
are known to be different. Namely, we need to consider general
three-flavor oscillations which require 9 real variables to fully
describe our issue. However, if $\theta_{13}$ is zero, a
simplification to two-flavor oscillations is possible by using
non-mixing mass eigenstates $\nu^\prime_\mu$ and $\nu^\prime_\tau$
instead of $\nu_\mu$ and $\nu_\tau$~\footnote{Similar simplification
is shown to be useful to analyze the evolution of
neutrino-antineutrino asymmetries by
Refs.~\cite{Abazajian:2002qx,Dolgov:2004jw,Wong:2002fa}. }.  Then $\nu^\prime_\mu$
and $\nu_e$ are described by two-flavor oscillations and
$\nu^\prime_\tau$ decouples from the oscillations. $\nu^\prime_\tau$
just interacts with $e^\pm$ via neutral current and should behave as
$\nu_\tau$ (or $\nu_\mu$) in the no-mixing case.

Under those assumptions, the variables necessary for simulating
thermalization of oscillating neutrinos are: the inflaton energy
density $\rho_\phi$, the photon temperature $T$, the
$\nu_e$-$\nu^\prime_\mu$ two-flavor neutrino density matrix $\rho_p$
and the $\nu^\prime_\tau$ distribution function
$f_{\nu^\prime_\tau}(p)$. $\rho_p$ and $f_{\nu^\prime_\tau}$ are
functions of neutrino momentum $p$. $\rho_p$ is defined by expectation
value of the product of the creation and annihilation
operators~\cite{Sigl:1992fn}:
\bea
\la a_j^\dagger({\bf p})a_i({\bf q}) \ra& \equiv &
 \left(2 \pi \right)^3 \delta^{(3)}\left({\bf p}-{\bf q}\right)
\left[\rho_p \right]_{ij}, ~~~\{ i,j \} =\{e,\mu\},
\eea
where $a_i({\bf p})$ is the annihilation operator for
negative-helicity neutrino of flavor $i$ with momentum ${\bf p}$.
Readers should bear in mind that the density matrix $\rho_p$ is just
an extension of the occupation number to the mixed neutrinos, and
should not confuse with the energy density, to which we refer as
$\rho_\nu$, $\rho_\phi$, etc.  Each diagonal component of $\rho_p$ is
the neutrino distribution of the corresponding flavor, while the
off-diagonal ones represent more subtle information on the
correlation.  For anti-neutrinos we can similarly define the density
matrix $\bar{\rho}_p$:
\bea
\la b_i^\dagger({\bf p})b_j({\bf q}) \ra& \equiv &
 \left(2 \pi \right)^3 \delta^{(3)}\left({\bf p}-{\bf q}\right)\left[\bar\rho_p \right]_{ij}, ~~~\{ i,j \} =\{e,\mu\},
\eea
where $b_i({\bf p})$ is the annihilation operator for
positive-helicity neutrino of flavor $i$ with momentum ${\bf p}$.
However, unless the lepton asymmetry is very large, we do not have to
distinguish neutrinos from anti-neutrinos. In this case they are
related to each other as $\bar \rho_p = \rho_p^T$.  We next derive the
differential equations which govern their evolutions.

We use scale factor $a$ as a time variable and later we use $y \equiv
p a$ instead of a momentum \cite{Dolgov:1997mb}. Then the time
evolution equation for the neutrino density matrix $\rho_p$
is~\cite{Sigl:1992fn}
\begin{eqnarray}
H a \frac{d \rho_p}{d a} =  - i [ \Omega(p), \rho_p] + I_{\rm coll}(p). 
\label{eq:diff_Fnu}
\end{eqnarray}
The matrix $ \Omega(p)$ represents both the vacuum oscillations and
the refractive term. Neglecting lepton asymmetry, which is usually as
small as the baryon asymmetry, it is written as
\begin{eqnarray}
\label{eq:refractive}
\Omega(p) &\equiv& \Omega_V(p) - \frac{8\sqrt{2} G_F p}{3 m_W^2} E, 
\end{eqnarray}
where the Fermi coupling constant $G_F=1.16637 \times 10^{-11}$
MeV$^{-2}$, $W$ boson mass $m_W = 80 $ GeV.  In the ultrarelativistic
limit, $\Omega_V(p)$ is given by
\begin{eqnarray}
\Omega_V(p) = \frac{1}{2p}U M^2 U^T,
 \end{eqnarray}
where $M^2$, the neutrino mass matrix, and $U$, the matrix which
relates mass eigenstates and flavor eigenstates, are
\begin{eqnarray}
M^2 \equiv  \left(
\begin{array}{cc}
m_1^2 & 0 \\
0 & m_2^2 \\
\end{array}
\right),~~~~~
U \equiv  \left(
\begin{array}{cc}
\cos \theta_{12} & \sin \theta_{12} \\
-\sin \theta_{12} & \cos \theta_{12} \\
\end{array}
\right).
\end{eqnarray}
We use the solar neutrino oscillation experiment values for neutrino
parameters: $m_2^2-m_1^2 = 7.3 \times 10^{-5}$ eV$^2$ and $\sin^2
\theta_{12} = 0.315$~\cite{Fogli:2003kp}~\footnote{
The most recent result~\cite{solar_nu} gives a slightly higher value
of the mass squared difference: $m_2^2-m_1^2= 7.9^{+0.6}_{-0.5} \times
10^{-5}$ eV$^2$.  However, we have confirmed that our results do not
change for $m_2^2-m_1^2$ in the error range.  }. The second term in
\EQ{eq:refractive} comes from the non-local effect of the $W$-exchange
interactions, and $E$ is the energy density matrix of the charged
leptons:
\begin{eqnarray}
E = 
\left(
\begin{array}{cc}
\rho_e+\rho_{\bar e} & 0 \\
0 & 0 \\
\end{array}
\right) =
 \left(
\begin{array}{cc}
(7/60)\pi^2 T^4 & 0 \\
0 & 0 \\
\end{array}
\right),
\end{eqnarray}
where $\rho_{e(\bar e)}$ is the energy density of electrons
(positrons) and we have assumed that neither muons nor taus exist in
the plasma.

For the collision term $I_{\rm coll}$, we consider the processes $\nu
+ e^\pm \leftrightarrow \nu + e^\pm$ and $\nu + \bar{\nu}
\leftrightarrow e^- + e^+$. In calculating the collision term, 
we take electrons to be massless and
neglect processes of scattering among neutrinos as
Refs.~\cite{Kawasaki:1999na, Kawasaki:2000en, Hannestad:2004px}. The
contributions from each process are
\begin{eqnarray}
I_{\nu e\nu e}(p_1) &=&
 \frac{1}{2 E_1} \int \frac{d {\bf p}_2}{2 E_2} \frac{d {\bf p}_3}{2 E_3}
\frac{d {\bf p}_4}{2 E_4} \, (2 \pi)^4 \delta^{(4)}(p_1+p_2-p_3-p_4)\,
\nonumber \\
\times&2^5 G_F^2 &\left[4 (p_1\cdot p_2) (p_3 \cdot p_4)\,
 F_{LL}(\nu^{(1)},e^{(2)},\nu^{(3)},e^{(4)})\right. \nonumber \\
&&\left. +4 (p_1\cdot p_4) (p_2 \cdot p_3)\, 
F_{RR}(\nu^{(1)},e^{(2)},\nu^{(3)},e^{(4)})  \right], \\
I_{\nu \bar e\nu \bar e}(p_1) 
&=& \frac{1}{2 E_1} \int \frac{d {\bf p}_2}{2 E_2} \frac{d {\bf p}_3}{2 E_3}
\frac{d {\bf p}_4}{2 E_4} \, (2 \pi)^4 \delta^{(4)}(p_1+p_2-p_3-p_4)\,
\nonumber \\
\times&2^5 G_F^2 &\left[4 (p_1\cdot p_4) (p_2 \cdot p_3)\, 
F_{LL}(\nu^{(1)},\bar e^{(2)},\nu^{(3)},\bar e^{(4)})\right. \nonumber \\
&&\left. +4 (p_1\cdot p_2) (p_3 \cdot p_4)\, 
F_{RR}(\nu^{(1)},\bar e^{(2)},\nu^{(3)},\bar e^{(4)}) \right], \\
I_{\nu \bar \nu e \bar e}(p_1)
&=& \frac{1}{2 E_1} \int \frac{d {\bf p}_2}{2 E_2} \frac{d {\bf p}_3}{2 E_3}
\frac{d {\bf p}_4}{2 E_4} \, (2 \pi)^4 \delta^{(4)}(p_1+p_2-p_3-p_4)\,
\nonumber \\
\times&2^5 G_F^2 &\left[4 (p_1\cdot p_4) (p_2 \cdot p_3)\, 
F_{LL}(\nu^{(1)},\bar \nu^{(2)},e^{(3)},\bar e^{(4)})\right. \nonumber \\
&&\left. +4 (p_1\cdot p_3) (p_2 \cdot p_4)\,
 F_{RR}(\nu^{(1)},\bar \nu^{(2)},e^{(3)},\bar e^{(4)}) \right],
\end{eqnarray}
where we define $d {\bf p} \equiv d^3 {\bf p}/(2\pi)^3$, $E_i \equiv p_i^0$, and
\begin{eqnarray}
F_{ab}(\nu^{(1)},e^{(2)},\nu^{(3)},e^{(4)})
&\equiv&\frac{1}{2} \left[(1-\rho_{p_1}) G_a
\rho_{p_3} G_b\,(1-f_e(p_2))f_e(p_4) + {\rm h.c.} \right. \nonumber\\
&&~~~~~~-\left. \rho_{p_1} G_a (1- \rho_{p_3}) G_b f_e(p_2)
(1-f_e(p_4)) + {\rm h.c.} \right], \\
F_{ab}(\nu^{(1)},\bar e^{(2)},\nu^{(3)},\bar e^{(4)})
&\equiv&\frac{1}{2} \left[(1-\rho_{p_1}) G_a
\rho_{p_3} G_b\,(1-f_{\bar e}(p_2)) f_{\bar e}(p_4)+ {\rm h.c.} \right. \nonumber\\
&&~~~~~~\left.-\rho_{p_1} G_a (1-\rho_{p_3}) G_b f_{\bar e}(p_2)
(1-f_{\bar e}(p_4))+ {\rm h.c.}\right] ,\\
F_{ab}(\nu^{(1)},\bar \nu^{(2)},e^{(3)},\bar e^{(4)})
&\equiv&\frac{1}{2} \left[(1-\rho_{p_1}) G_a
(1-\bar\rho_{p_2}) G_b\,f_e(p_3) f_{\bar e}(p_4)+ {\rm h.c.} \right. \nonumber\\
&&~~~~~~\left.-\rho_{p_1} G_a \bar\rho_{p_2} G_b  (1-f_e(p_3))
(1-f_{\bar e}(p_4))+ {\rm h.c.}\right] ,
\end{eqnarray}
with $G_L = {\rm diag} (g_L,\,\tilde g_L)$ and $G_R= {\rm diag}(g_R,\,
g_R)$. Here, $\tilde{g}_L=g_L-1 = \sin^2 \theta_W-\frac{1}{2}$ and
$g_R = \sin^2 \theta_W$ where $\sin^2 \theta_W = 0.23120$ is the
weak-mixing angle. $f_{e} (f_{\bar e} )$ is the distribution function
of electrons (positrons).  Hereafter we take $\bar \rho_p = \rho_p^T$
and $f_{\bar e} = f_e$. Note that these collision terms coincide with
those found in Ref.~\cite{Dolgov:1997mb} if oscillations are absent
(i.e., if off-diagonal components in the density matrices are zero).

We further approximate that electrons obey the Boltzmann distribution
and their Pauli blocking factors are neglected. Namely, in $F$'s, we
replace as $f_e(p) \rightarrow \exp(-p/T)$ and $1-f_e(p) \rightarrow
1$. Then the collision terms above are reduced to one-dimensional
momentum integration by the technique in Ref.~\cite{Kawasaki:2000en}
and the reduced expressions become equal to the ones in the
reference~\footnote{The right hand side of Eq.~(A16) in
Ref.~\cite{Kawasaki:2000en} has to be multiplied by 2 to be a correct
equation. Due to this error, the right hand side of Eq.~(3) in
Ref.~\cite{Kawasaki:1999na} has to be multiplied by 2. The right hand
side of Eq.~(12) in Ref.~\cite{Kawasaki:2000en} has to be multiplied
by 8 since it had already contained a typo of factor 4. In this
occasion, we correct a typo in the right hand side of Eq.~(8) in
Ref.~\cite{Kawasaki:2000en}: it has to multiplied by 2 (so that it is
same as Eq.~(2) in Ref.~\cite{Kawasaki:1999na}.}  in the limit of the
zero mixing angle.

In practice, since $\rho_p$ is 2$\times$2 Hermitian matrix, it is
convenient to expand it using Pauli matrices. Namely,
\begin{eqnarray}
\rho_p = \sum_{i=0}^3 P_i(p) \frac{\sigma_i}{2},
\end{eqnarray}
where
\begin{eqnarray}
\sigma_0 \equiv  \left(
\begin{array}{cc}
1 & 0 \\
0 & 1 \\
\end{array}
\right),  \quad
\sigma_1 \equiv  \left(
\begin{array}{cc}
0 & 1 \\
1 & 0 \\
\end{array}
\right),  \quad
\sigma_2 \equiv  \left(
\begin{array}{cc}
0 & -i \\
i & 0 \\
\end{array}
\right),  \quad
\sigma_3 \equiv  \left(
\begin{array}{cc}
1 & 0 \\
0& -1 \\
\end{array}
\right).
\end{eqnarray}
On the right hand side of Eq.~(\ref{eq:diff_Fnu}), $i[\Omega, \rho_p]$
and $I_{\rm coll}$, are expanded similarly. We solve for the evolution
of $P_0 \sim P_3$ and the distributions of $\nu_e$ and
$\nu^\prime_\mu$ are in turn derived by $f_{\nu_e} = (P_0+P_3)/2$ and
$f_{\nu^\prime_\mu} = (P_0-P_3)/2$. The evolution equations are
formally written as
\begin{eqnarray}
H a \frac{d P_i(y)}{da} = -i \Omega_i(y) + I_i(y), \label{eq:diff_f}
\end{eqnarray}
where $i$ runs from 0 to 3, $\Omega_i \equiv {\rm Tr}([\Omega, \rho_p
]\sigma_i)$ and $I_i \equiv {\rm Tr}(I_{\rm coll} \,\sigma_i)$, and we
have changed the variable $p$ to $y$.

We need to solve for the evolution of $\nu^\prime_\tau$, too. To this
end, it is most simple to obtain the time evolution of
$f_{\nu^\prime_\tau}$ from the $\nu^\prime_\mu$-component of
Eq.~(\ref{eq:diff_Fnu}) with no mixing (which is given by omitting the
first term on the right hand side) because the interactions of
$\nu^\prime_\tau$ with $e^\pm$ are same as those of $\nu^\prime_\mu$.

For the evolution of $\rho_\phi$ and $T$, the equations are almost
same as those found in Ref.~\cite{Kawasaki:2000en}. We just need
modifications due to our use of scale factor $a$ as a time variable
and discrimination of $\nu_\mu$ from $\nu_\tau$. For $\rho_\phi$, it
is given by
\begin{eqnarray}
\frac{d\rho_\phi}{da} = -\frac{\Gamma}{aH} \rho_\phi - \frac{3}{a}\rho_\phi. \label{eq:diff_phi}
\end{eqnarray}
The equation of the total energy-momentum conservation is
\begin{eqnarray}
\label{eq:ev_tot}
\frac{d\rho_{\rm tot}}{da} =  - \frac{3}{a}(\rho_{\rm tot} + P_{\rm tot}),
\end{eqnarray}
where the total energy density and the total pressure are given by
\bea
\rho_{\rm tot} & \equiv & \rho_\phi +  \rho_\gamma + \rho_{e^\pm} 
+ \rho_{\nu_e} + \rho_{\nu^\prime_\mu}+ \rho_{\nu^\prime_\tau}\non\\
&=& \rho_\phi + \frac{\pi^2 T^4}{15}
+ \frac{2}{\pi^2} \int_0^\infty dp  \,p^2 \frac{E_e}{\exp(E_e/T)+1}
\non\\&&
+ \frac{1}{\pi^2 a^4 }\int _0^\infty dy\,y^3 \left (f_{\nu_e}+ f_{\nu^\prime_\mu}+f_{\nu^\prime_\tau}\right),\\
P_{\rm tot} &\equiv& P_\gamma + P_{e^\pm} 
+ P_{\nu_e} + P_{\nu^\prime_\mu}+ P_{\nu^\prime_\tau}, \non\\
&=& \frac{\pi^2 T^4}{45}
+ \frac{2}{3\pi^2} \int_0^\infty dp \frac{p^4}{E_e \,\left(\exp(E_e/T)+1\right)} \non\\
&&+ \frac{1}{3 \pi^2 a^4 }\int _0^\infty dy\,y^3 \left (f_{\nu_e}+ f_{\nu^\prime_\mu}+f_{\nu^\prime_\tau}\right)
\eea
with the electron energy $E_e \equiv \sqrt{m_e^2 + p^2}$.
The evolution equation for $T$ is derived from \EQ{eq:ev_tot}:
\bea
\frac{dT}{da} &=& - \left( \frac{\partial \rho_\gamma}{\partial T} 
	+\frac{\partial \rho_{e^\pm}}{\partial T}\right)^{-1} \left\{ \frac{4}{a}\rho_\gamma 
	+\frac{3}{a}(\rho_{e^\pm}+P_{e^\pm}) -\frac{\Gamma}{aH}  \rho_\phi \right. \non\\
&&\left.	+ \frac{1}{\pi^2 a^4} \int _0^\infty dy\,y^3 \left (\frac{df_{\nu_e}}{da} 
	+ \frac{df_{\nu^\prime_\mu}}{da} +\frac{df_{\nu^\prime_\tau}}{da} \right)  \right\}
	\label{eq:diff_temp}
\eea
Finally, the expansion rate is 
\begin{eqnarray}
H \equiv \frac{1}{a} \frac{da}{dt} = \frac{\sqrt{\rho_{\rm tot}} }{\sqrt{3} M_{\rm Pl}}.
\end{eqnarray}

To integrate the differential equations, since the equations for
$f_\nu(y)$ are stiff, we used semi-implicit extrapolation method
\cite{NumericalRecipes}. Using the Ref.~\cite{NumericalRecipes}'s
implementation which incorporates adaptive stepsize control routine,
we were able to evolve the neutrino density matrices very
efficiently. We followed the evolution well after the
electron-positron annihilation ends and $f_\nu(y)$'s become constant.

As for the initial condition, we have to make the inflaton energy
density dominate the universe at first. As long as $\rho_\phi$ is much
larger than radiation energy density ($\sim T^4$), evolution afterward
does not depend on their precise values. In this paper, we adopt
rather realistic relation between $\rho_\phi$ and $\rho_{\rm rad}$,
\begin{eqnarray}
\rho_{\rm rad} &=& \frac{2\sqrt{3}}{5} \Gamma M_{\rm Pl} \rho_\phi^{1/2},
\end{eqnarray}
which derived from the analytic solutions during the epoch of coherent
oscillations \cite{EarlyUniverse}.

\section{Results and cosmological implications} \label{sec:implications}
 In this section, we present the results of our numerical calculation
 for neutrino thermalization and consider its implications for
 cosmology. We evolve the neutrino density matrices with various
 values of the reheating temperature $T_R$ and investigate how the
 neutrino distribution functions, neutrino energy densities and big
 bang nucleosynthesis depend on $T_R$. Along with the neutrino
 thermalization with oscillations, we show the results without
 oscillations which have been studied in Refs~\cite{Kawasaki:1999na,
 Kawasaki:2000en, Hannestad:2004px} and elucidate the neutrino
 oscillation effects on a low reheating temperature scenario. Our
 results when the oscillations are omitted turn out to be consistent
 with those of previous papers. We find that the inclusion of the
 oscillations most characteristically alters the $^4$He synthesis and
 its abundance varies with respect to $T_R$ quite differently from the
 no oscillation case .

\subsection{Neutrino distribution functions} \label{sec:fnu}
\begin{figure}
\begin{center}
\begin{tabular}{cc}	
\includegraphics[width=7.5cm]{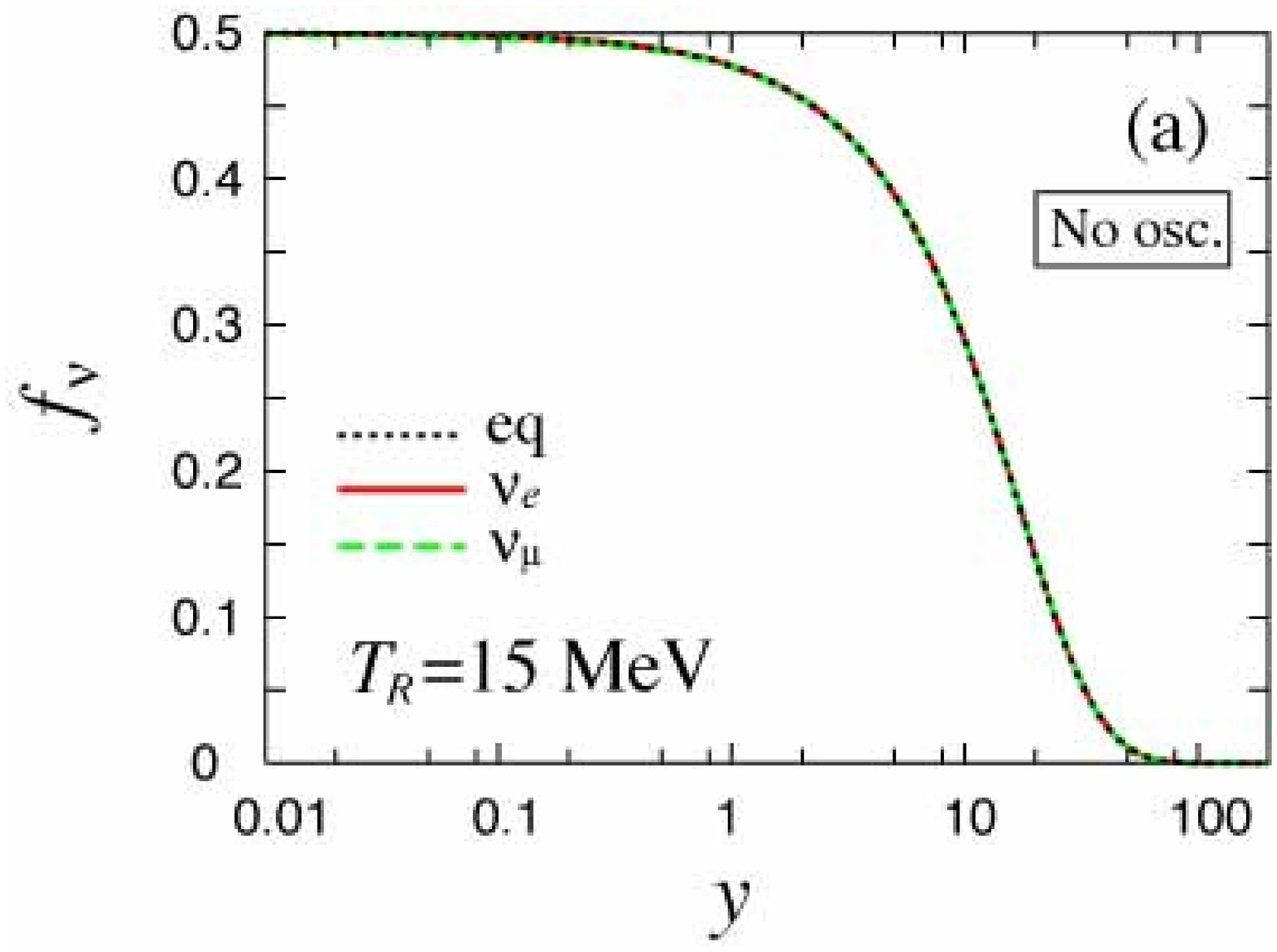} &
\includegraphics[width=7.5cm]{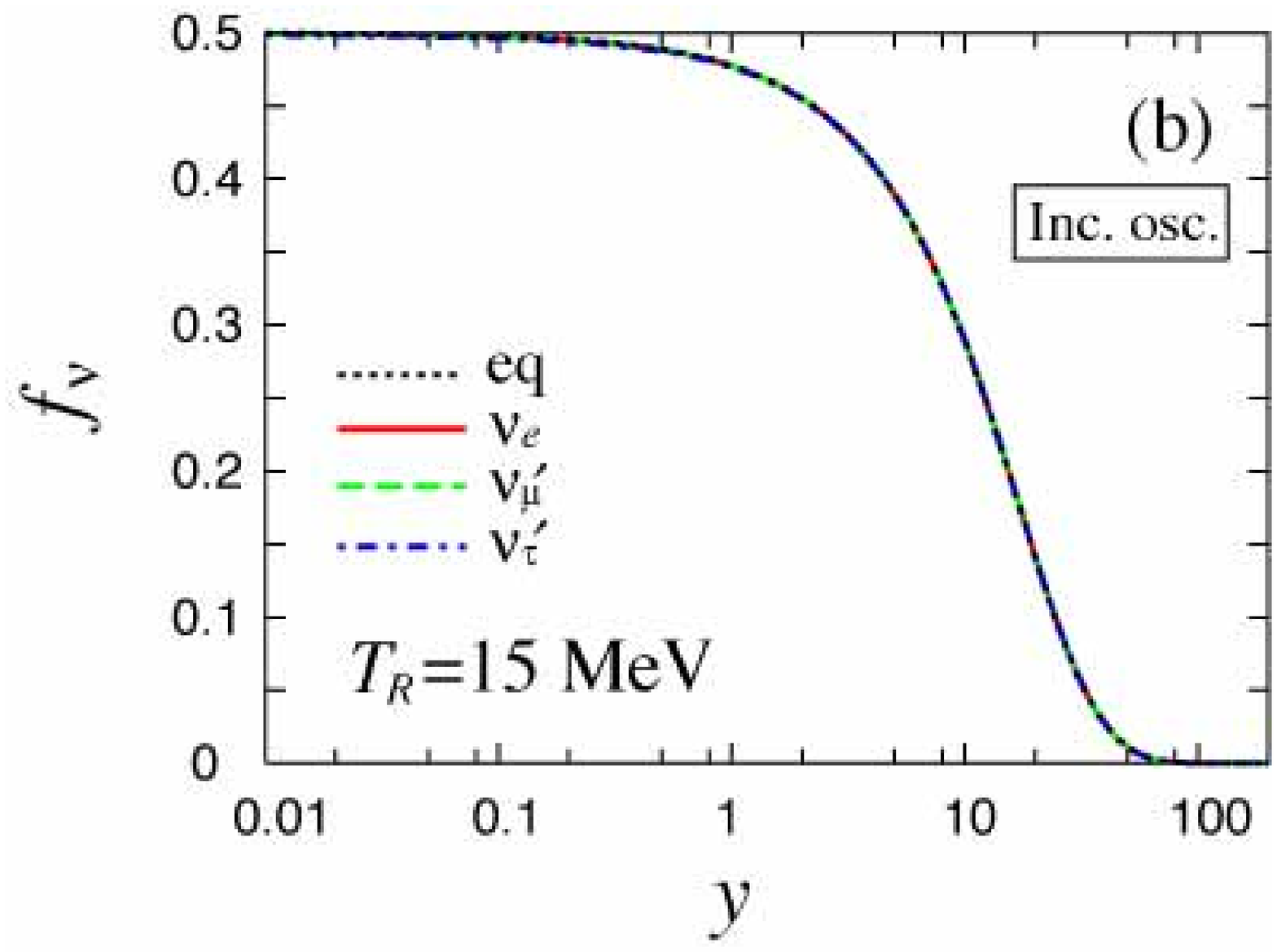}\\
\includegraphics[width=7.5cm]{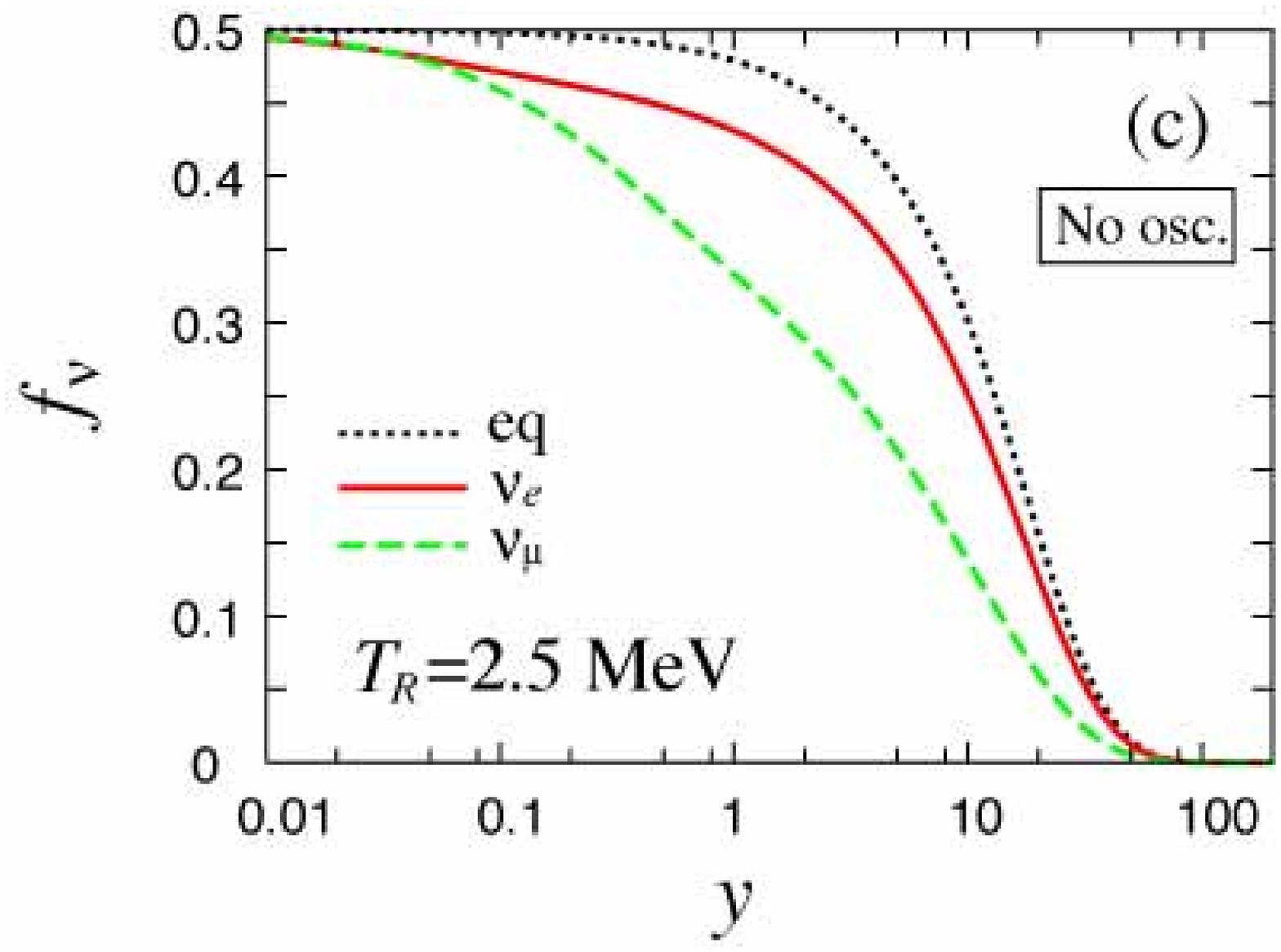}  &
\includegraphics[width=7.5cm]{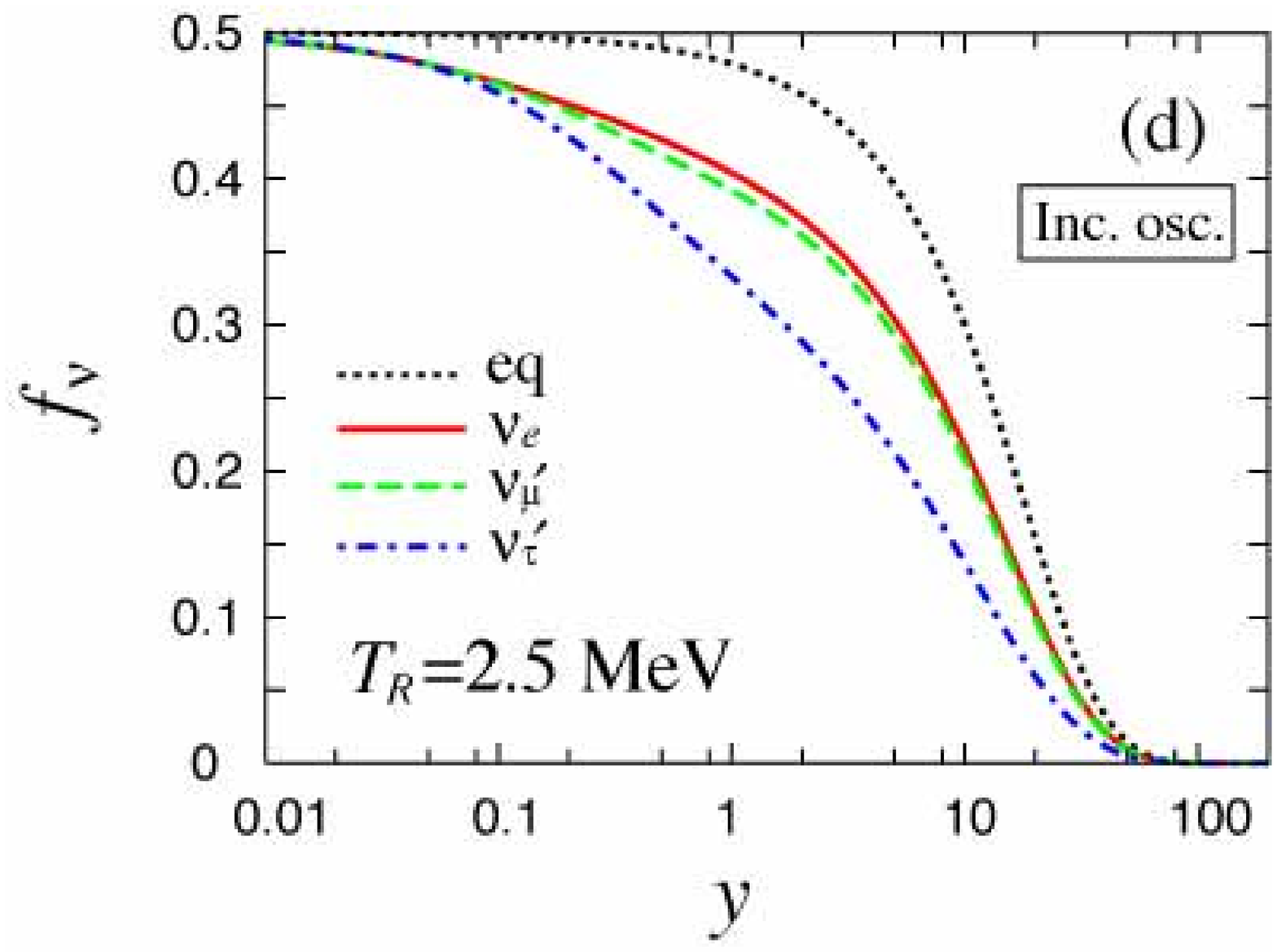} \\
\end{tabular}
\caption{
The final distribution functions of neutrinos. (a) and (c) are cases
for no oscillations ($\nu_e$ is displayed by solid lines and $\nu_\mu$
by dashed lines) and (b) and (d) incorporate the oscillations ($\nu_e$
is displayed by solid lines, $\nu^\prime_\mu$ by dashed lines and
$\nu^\prime_\tau$ by dot-dashed lines). The equilibrium distributions
are drawn by dotted lines in order to show how much they are
thermalized. For $T_R=15$ MeV, in (a) and (b), whether the
oscillations are present or not, all the lines overlap and this means
every neutrino species is fully thermalized for high reheating
temperature. For $T_R=2.5$ MeV, in (c) and (d), distributions are away
from equilibrium form. When the oscillations are taken into account,
distributions of $\nu_e$ and $\nu^\prime_\mu$ get close as seen in
(d).
}
\label{fig:dist}
\end{center}
\end{figure}

We show the final neutrino distribution functions in
Figs.~\ref{fig:dist} (a)-(d) for the cases of $T_R=$ 15 MeV and 2.5
MeV respectively with and without the oscillations. We see from
Figs.~\ref{fig:dist} (a) and (b) that when the reheating temperature is
sufficiently high, all the neutrino species are thermalized regardless
of the oscillations. 

\begin{figure}
\begin{center}
\begin{tabular}{cc}	
\includegraphics[width=12cm]{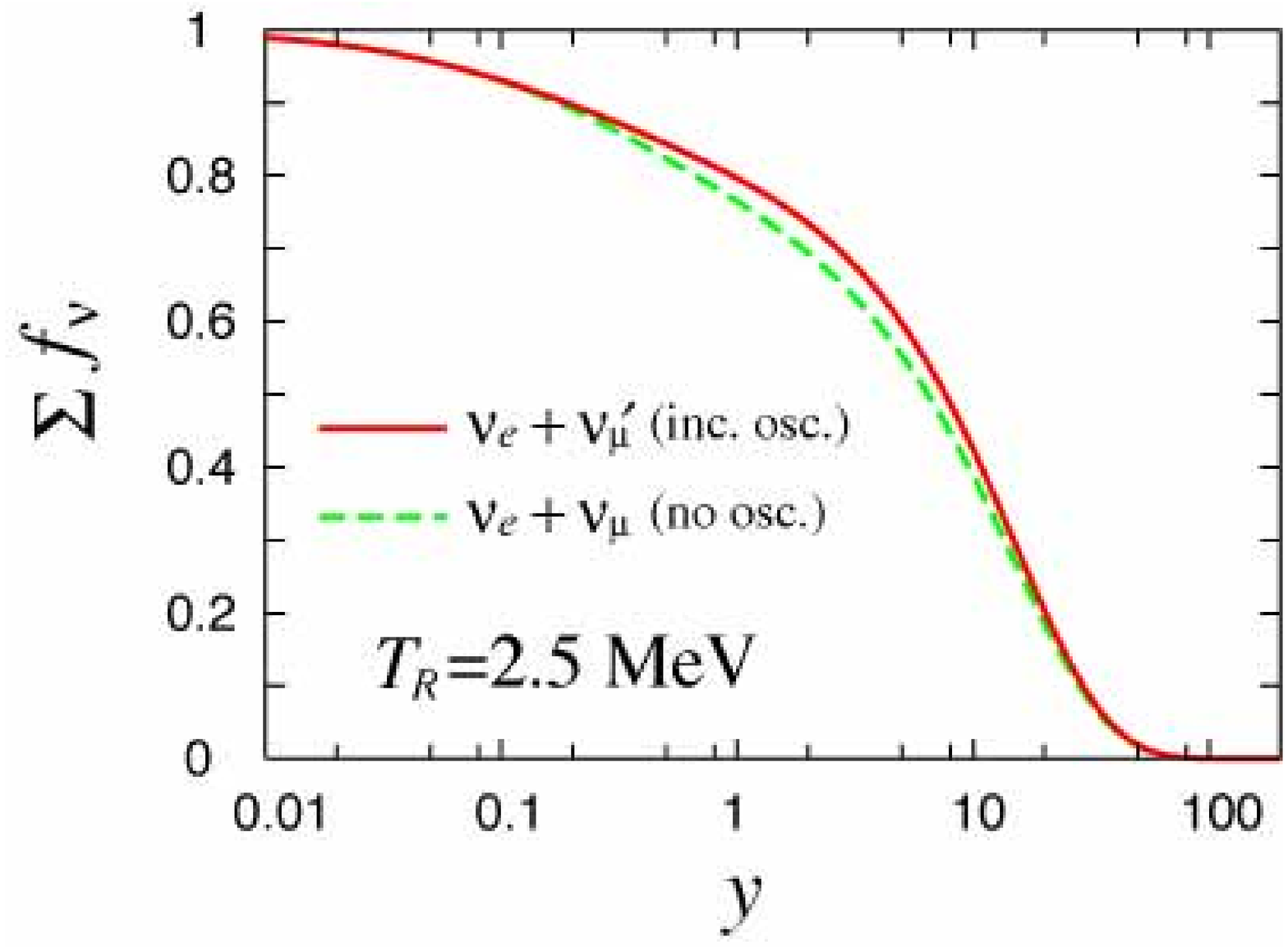} 
\end{tabular}
\caption{
We draw the
sums of the distribution functions, $f_{\nu_e}+f_{\nu_\mu}$ (no
oscillation) and $f_{\nu_e}+f_{\nu^\prime_\mu}$ (including
oscillation) with the dashed line and the solid line respectively. The
latter is larger showing that the oscillations make the thermalization
more efficient in total.
}
\label{fig:dist_compare}
\end{center}
\end{figure}

However, for the case of lower reheating temperature, the oscillations
significantly matter as seen from comparing Figs.~\ref{fig:dist} (c)
and (d) : $f_{\nu_e}$ and $f_{\nu^\prime_\mu}$ are almost equalized by
the solar mixing. When the oscillations are neglected, $f_{\nu_e}$
becomes much larger than $f_{\nu_\mu}$ as shown in Fig.~\ref{fig:dist}
(c) because $\nu_e$ is produced by the charged current interaction in
addition to the neutral current interaction but $\nu_\mu$ and
$\nu_\tau$ are produced only by the latter
\cite{Kawasaki:1999na, Kawasaki:2000en}. When there are the
flavor mixings, $\nu_e$ and $\nu^\prime_\mu$ can convert into each
other. $\nu^\prime_\mu$ is now produced also by the oscillations from
$\nu_e$ which exists more than $\nu^\prime_\mu$ so
$f_{\nu^\prime_\mu}$ increases compared to no oscillation case. On the
contrary, $f_{\nu_e}$ becomes smaller when the oscillations are
included naturally because $\nu_e$ oscillates into
$\nu^\prime_\mu$. However, this deficit is to some extent filled by
the $\nu_e$ production from the thermal plasma so the neutrinos are
produced more in total under the existence of the oscillations. This
is seen in Fig.~\ref{fig:dist_compare} which shows clearly that
$f_{\nu_e}+f_{\nu^\prime_\mu}$ (including oscillation) $>$
$f_{\nu_e}+f_{\nu_\mu}$ (no oscillation).

\subsection{Effective number of neutrinos} \label{sec:Nnu}
Let us discuss our results of the neutrino thermalization in terms of
neutrino energy density. This is often expressed using the effective
number of neutrinos $N_\nu$. This number is observationally relevant
to the CMB power spectrum and large scale structure. It is given by
\begin{eqnarray}
N_\nu \equiv \frac{\sum \rho_\nu}{\rho_{\nu, {\rm std}}}, \label{eq:Nnu_def}
\end{eqnarray}
where the summation is taken for $\nu = \nu_e,\ \nu_\mu,$ and $ \nu_\tau$
when the oscillations are not included and $\nu = \nu_e,\
\nu^\prime_\mu,$ and $\nu^\prime_\tau$ when we consider the
oscillations. 
 We define $\rho_{\nu, {\rm std}}$ using the photon temperature
$T$ as
\begin{eqnarray}
\rho_{\nu, {\rm std}} = \frac{7 \pi^2}{120} 
	\left\{ \left(\frac{4}{11}\right)^{1/3} T \right\}^4,
\end{eqnarray}
which corresponds to the neutrino energy density assuming that
neutrinos are completely decoupled from the rest of the thermal plasma
before the electron-positron annihilation takes place. If this
assumption is exact, $N_\nu$ would be 3. It is actually a very good
assumption but detailed calculations on the entropy transfer from
electrons to neutrinos have shown that $\rho_\nu$'s are slightly
larger than $\rho_{\nu, {\rm std}}$ and $N_\nu = 3.04$
\cite{Hannestad:1995rs,Dolgov:1997mb,Dolgov:1998sf,Mangano:2001iu}.

\begin{figure}
\begin{center}
\includegraphics[width=12cm]{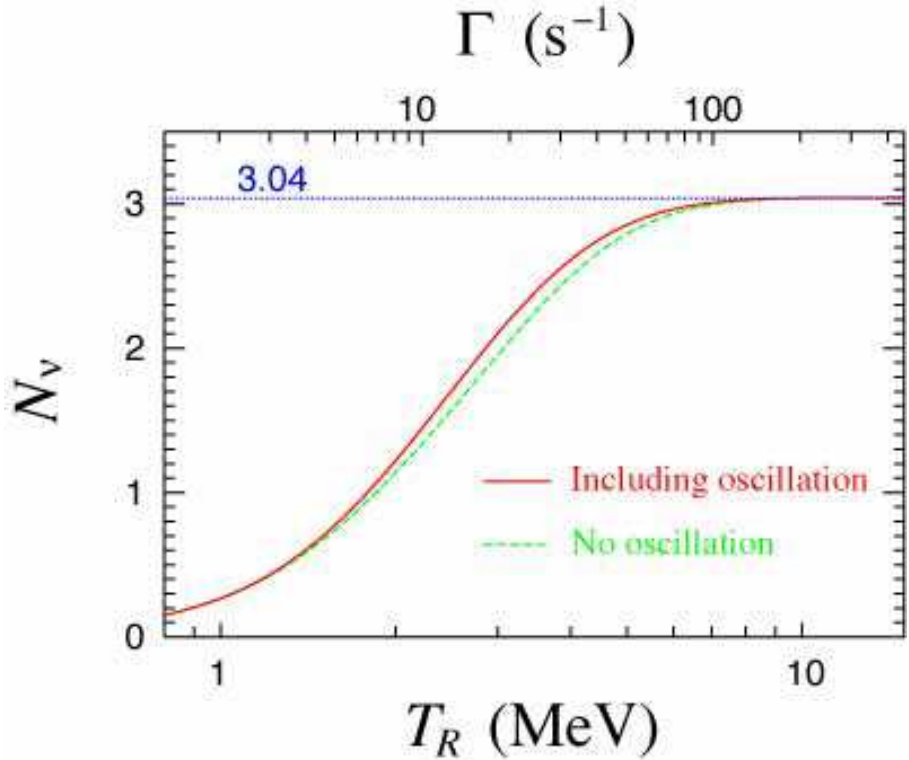} 
\caption{
The effective neutrino number $N_\nu$ as a function of the reheating
temperature $T_R$ (shown on the bottom abscissa) or the decay width
$\Gamma$ (shown on the top abscissa). The cases with and without the
oscillations are drawn respectively by the solid and dashed lines. The
horizontal line denotes $N_\nu=3.04$ with which $N_\nu$ for high $T_R$
should coincide (see the text).
}
\label{fig:Nnu_compare}
\end{center}
\end{figure}

We calculate $\rho_\nu$ by integrating the final neutrino distribution
functions such as presented in \FIG{fig:dist}, and derive $N_\nu$ as a
function of the reheating temperature $T_R$. The result is shown in
Fig.~\ref{fig:Nnu_compare}. For $T_R \gtrsim 10$ MeV, $N_\nu$
asymptotes the value 3.04 which indicates thermalized neutrino
distributions. This is regardless of the neutrino oscillations and
consistent with Fig.~\ref{fig:dist} (a) and (b) discussed in
section~\ref{sec:fnu}. For the smaller values of $T_R$, the inclusion
of the oscillations make $N_\nu$ larger as expected from
Fig.~\ref{fig:dist_compare}. This effect is most conspicuous for $T_R
= 2 \sim 5$ MeV and changes $N_\nu$ up to $\sim 0.2$.

Fig.~\ref{fig:Nnu_compare} enables us to constrain $T_R$ by using the
limits on the effective number of neutrino species from cosmological
observations such as CMB and galaxy surveys. Recent papars,
Refs.~\cite{Crotty:2003th,Pierpaoli:2003kw,Hannestad:2003xv,Barger:2003zg},
derive the lower limit to be $0.9 \sim 1.9$ (these are the limits
obtained without resorting to BBN. Some of them have also reported
more stringent limits obtained using data combined with observed
$Y_p$.  However, since they assume Fermi-Dirac distribution for
neutrinos and only modify the Friedmann equation when they calculate
$Y_p$, we cannot use those limits. This point is discussed in section
\ref{sec:bbn} in detail).  If $N_\nu > 0.9$ is adopted, the bound on
the reheating temperature is $T_R > 1.69$ MeV with the oscillations
and $T_R> 1.74$ MeV for no oscillation case.

\subsection{Light element abundances} \label{sec:bbn}
We now investigate how the big bang nucleosynthesis is affected by the
non-thermal neutrino distributions and/or the neutrino
oscillations. We calculate the light element (D, $^4$He and $^7$Li )
abundances as functions of $T_R$, again with and without the neutrino
oscillations. The cosmological effects of incomplete neutrino
thermalization is most strikingly seen in $^4$He abundance since
electron-type neutrinos play a special role in determining the rate of
neutron-proton conversion during BBN. This has been already known from
the previous papers Refs.~\cite{Kawasaki:1999na, Kawasaki:2000en} in
which the oscillations are neglected, but we find that the neutrino
oscillations prominently matter in regard to the $T_R$-dependence of
$^4$He abundance.

\begin{figure}
\begin{center}
\includegraphics[width=12cm]{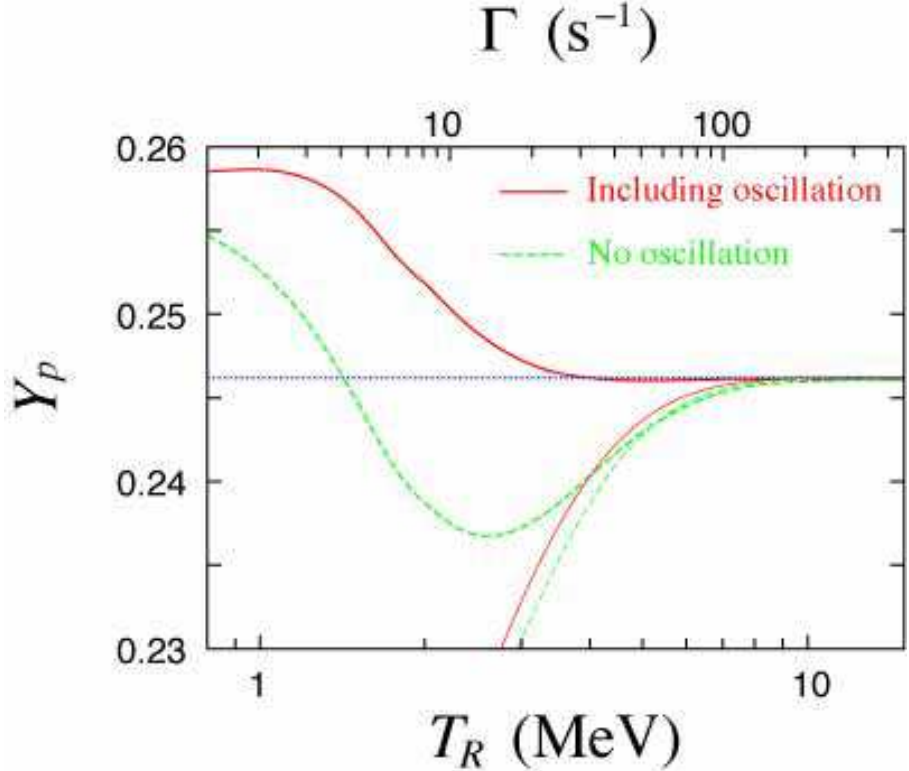} 
\caption{
The $^4$He abundance (mass fraction) $Y_p$ as a function of the
reheating temperature $T_R$ (shown on the bottom abscissa) or the
decay width $\Gamma$ (shown on the top abscissa). The cases with and
without the oscillations are drawn respectively by the solid and
dashed curves. Thinner curves are calculated with fermi distributed
neutrinos with $N_\nu$ of Fig.~\ref{fig:Nnu_compare} (namely, only the
change in the expansion rate due to the incomplete thermalization is
taken into account). The horizontal line represents ``standard " $Y_p$
calculated by BBN with neutrinos obeying the fermi distribution and
$N_\nu=3.04$. The baryon-to-photon ratio is fixed at $\eta = 5 \times
10^{-10}$.
}
\label{fig:TR_He4}
\end{center}
\end{figure}

We show how $Y_p$ varies with respect to $T_R$ in
Fig.~\ref{fig:TR_He4}. This is calculated by plugging the solutions of
the evolution equations derived in section~\ref{sec:thermalization} into
the Kawano BBN code \cite{Kawanocode} (with updated reaction rates
compiled by Angulo et al. \cite{NACRE}). Required modifications are
the temperature dependence of the  neutron-proton conversion rates, $\Gamma_{n
\rightarrow p}$ and $\Gamma_{p \rightarrow n}$, and the evolution
equation for the photon temperature. The calculation of $\Gamma_{n
\leftrightarrow p}$ (see e.g. Ref.~\cite{GraCos}) involves the
integration of the electron neutrino distribution function $f_{\nu_e}$
which does not necessarily take the Fermi distribution form in our
case. For the photon temperature evolution, the contributions from
$\phi$ and neutrinos are supplemented in the same way as
Eq.~(\ref{eq:diff_temp}).

There are two effects caused by incomplete thermalization of neutrinos
competing to make up the dependence of $Y_p$ on $T_R$ as shown in
Fig.~\ref{fig:TR_He4}: slowing down of the expansion rate and
decreasing in $\Gamma_{n \leftrightarrow p}$. The former is just a
result of the decrease in the neutrino energy density (of all
species). The latter is due to the deficit in $f_{\nu_e}$. They
compete in a sense that they work in opposite way to determine the
epoch of neutron-to-proton ratio freeze-out: the former makes it later
and the latter makes it earlier. Then, the competition fixes the n-p
ratio at the beginning of nucleosynthesis and eventually determine
$Y_p$. Roughly speaking, for larger $T_R$, the former dominates to
decreases $Y_p$ but, for smaller $T_R$, the latter dominates and
increase $Y_p$. This is clearly seen in the case without the
oscillations but not for the case including the oscillations because
the incompleteness in the $\nu_e$ thermalization is made severer by
the mixing (see the panels (c) and (d) in Fig.~\ref{fig:dist}) and
this effect dominates already at high $T_R$.

\begin{figure}
\begin{center}
\includegraphics[width=12cm]{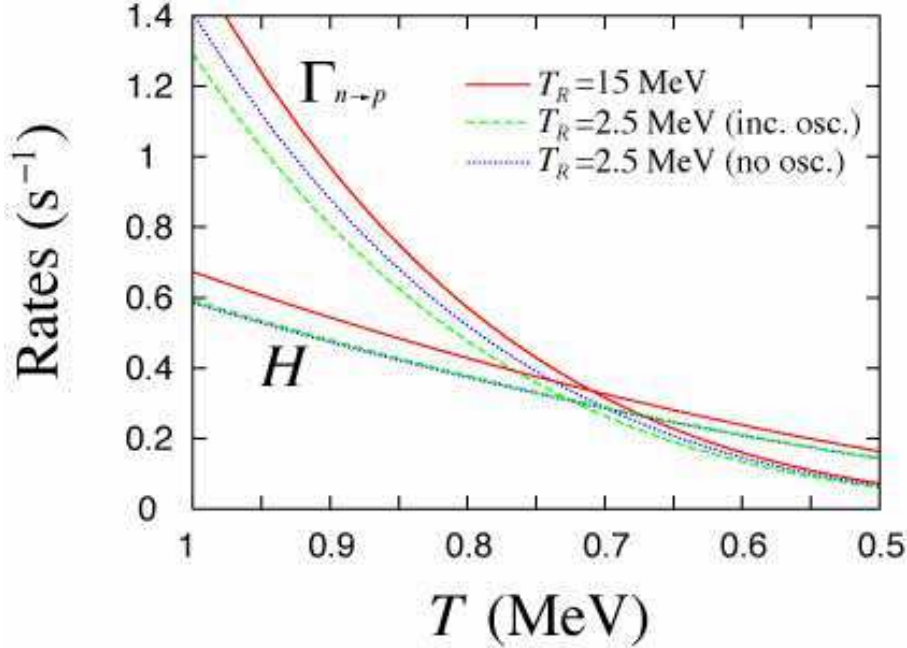} 
\caption{
The weak interaction rate $\Gamma_{n \rightarrow p}$ and the expansion
rate $H$ as functions of temperature, where
$\Gamma_{n \rightarrow p}$ and $H$ first become equal. We plot for $T_R = 2.5$ MeV with
and without the oscillations. For $T_R=15$ MeV, the oscillations do
not make any difference. 
}
\label{fig:Gamma_H}
\end{center}
\end{figure}

\begin{figure}
\begin{center}
\begin{tabular}{cc}	
\includegraphics[width=8cm]{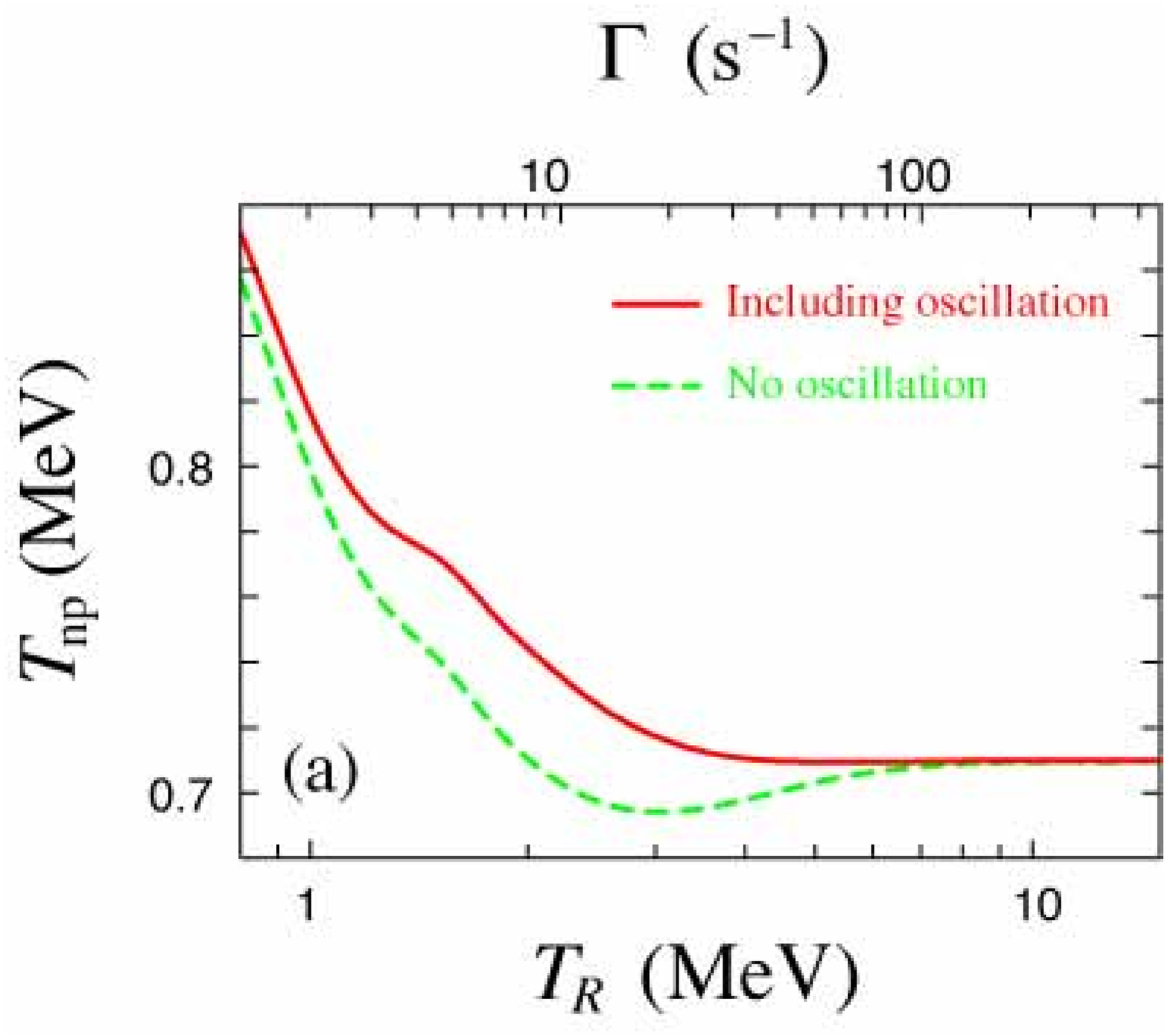} &
\includegraphics[width=8cm]{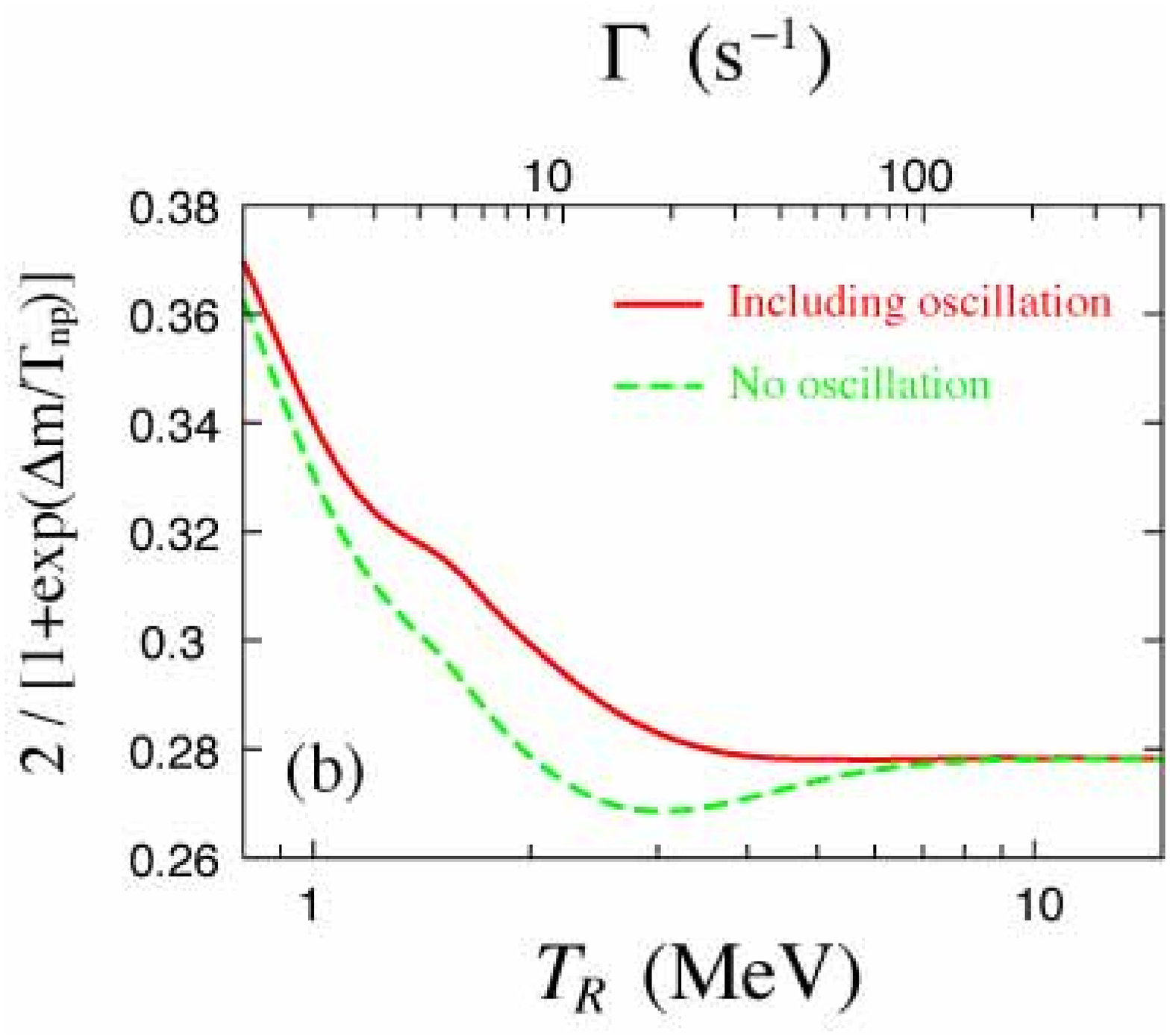} \\
\end{tabular}
\caption{
(a) $T_{\rm np}$, freeze-out temperature of the neutron-to-proton
ratio, and (b) $2/[1+\exp(\Delta m/T_{\rm np}) ]$, as functions of
$T_R$.
}
\label{fig:TR_Tnp}
\end{center}
\end{figure}

Before going forward, it may be worthwhile to look slightly more into
the explanation of the $T_R$-dependence of $Y_p$. First, let us forget
about modifying $\Gamma_{n \leftrightarrow p}$ or temperature
evolution and just calculate $^4$He abundance using thermally
distributed neutrinos with $N_\nu$'s indicated in
Fig.~\ref{fig:Nnu_compare} for each value of $T_R$. This corresponds
to including the effect of slowing down the expansion rate due to the
incomplete thermalization but neglecting the electron neutrino
deficiency. Accordingly, lowering $T_R$ only acts to delay the n-p
ratio freeze-out and decrease $Y_p$ (shown by the thinner curves in
Fig.~\ref{fig:TR_He4}). In actual low-reheating temperature scenario,
a lack of $\nu_e$ reduces $\Gamma_{n \leftrightarrow p}$. This
counterbalances the effect of slowing-down expansion and boosts $Y_p$
in total at lower $T_R$.  To see this is really the case, we plot
$\Gamma_{n \rightarrow p}$ for some values of $T_R$ in
Fig.~\ref{fig:Gamma_H}. We see that $\Gamma_{n \rightarrow p}$ is
smaller for lower $T_R$ which is attributed to less thermalized
$\nu_e$. It is also instructive to calculate the neutron-to-proton
ratio freeze-out temperature $T_{\rm np}$, which we define by
$\Gamma_{n \rightarrow p}(T_{\rm np}) = H(T_{\rm np})$, to confirm
where the competition settles. This is shown in Fig.~\ref{fig:TR_Tnp}
(a) and we see that at low $T_R$, the decrease in $\Gamma_{n
\rightarrow p}$ wins to make $T_{\rm np}$ higher (in the case with the
oscillations, this seems to win for every $T_R$ and $T_{\rm np}$ rises
monotonically as $T_R$ decreases). We note that the figure well
reproduces the profile found in Fig.~\ref{fig:TR_He4}. This
resemblance becomes more meaningful by plotting instead the quantity
$2/[1+(n_p/n_n)_f] = 2/[1+\exp(\Delta m/T_{\rm np}) ]$, the usual
estimation of $^4$He abundance from the neutron-to-proton ratio at the
freeze-out value, which is shown in Fig.~\ref{fig:TR_Tnp}
(b). Although the figure is not exactly same as Fig.~\ref{fig:TR_He4}
because free decays of neutrons are not considered, we see that the
$Y_p$'s dependence on $T_R$ is sufficiently understood from this
estimation. When the neutron free decay is properly taken into
account, the estimation for $Y_p$ decreases from the values indicated
in Fig.~\ref{fig:TR_Tnp} (b). For lower $T_R$, since the time between
$T_{\rm np}$ and the start of the nucleosynthesis ($T \approx 0.07$
MeV) is longer (this in turn is explained by the smaller expansion
rate due to less neutrino energy densities), this decrease should be
larger. Therefore, on including the neutron free decay,
Fig.~\ref{fig:TR_Tnp} (b) would be tilted toward left (smaller $T_R$)
side and should look more like Fig.~\ref{fig:TR_He4}. In particular,
the minimum found for the case without the oscillations should be
located at lower $T_R$ when the free decay is included.

We have so far discussed the $^4$He synthesis features common to the
low-$T_R$ universe with and without the neutrino oscillations, but we
would rather like to emphasize that there is a striking difference
between them. This is most clearly visible in Fig.~\ref{fig:TR_He4}:
when we include the oscillations, $Y_p$ does not decrease if we lower
$T_R$. This is somewhat surprising because, at the same time, $N_\nu$
becomes smaller (see Fig.~\ref{fig:Nnu_compare}). This means that, in
the case with the oscillations, the effect of slowing down cosmic
expansion (as represented by decreasing $N_\nu$) is completely
overcome by the decrease in $\Gamma_{n \leftrightarrow p}$ for all
$T_R$. The reason why this happens is that since the oscillations
convert electron neutrinos into muon neutrinos, the deficiency in
electron neutrinos is made severer (see
Fig.~\ref{fig:dist_compare}). Moreover, why this matters for $^4$He
synthesis is that it is exclusively sensitive to the $\nu_e$
distribution function which determines $\Gamma_{n
\leftrightarrow p}$. On the other hand, the structure formation is
affected only by the energy density so it does not distinguish
neutrino flavors. Since only their sum matters, the oscillations
scarcely make difference (see Fig.~\ref{fig:Nnu_compare}). Therefore,
BBN, especially when the neutrino oscillations are taken into account,
turns out to be unique probe of low reheating temperature scenario.
Next, we proceed to compare the predictions of the scenario with the
observed abundances.

\begin{figure}
\begin{center}
\begin{tabular}{cc}	
\includegraphics[width=7.5cm]{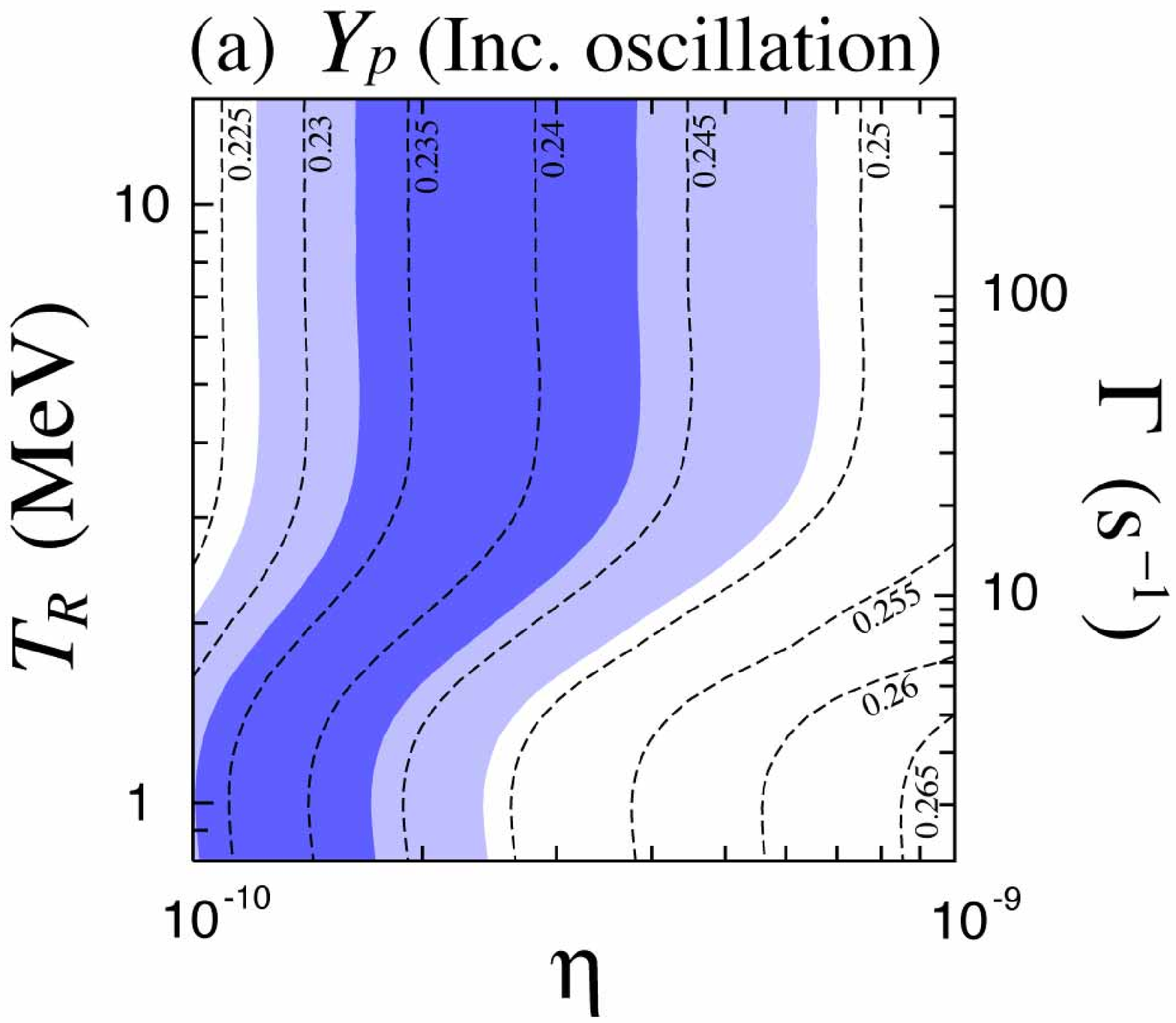} &
\includegraphics[width=7.5cm]{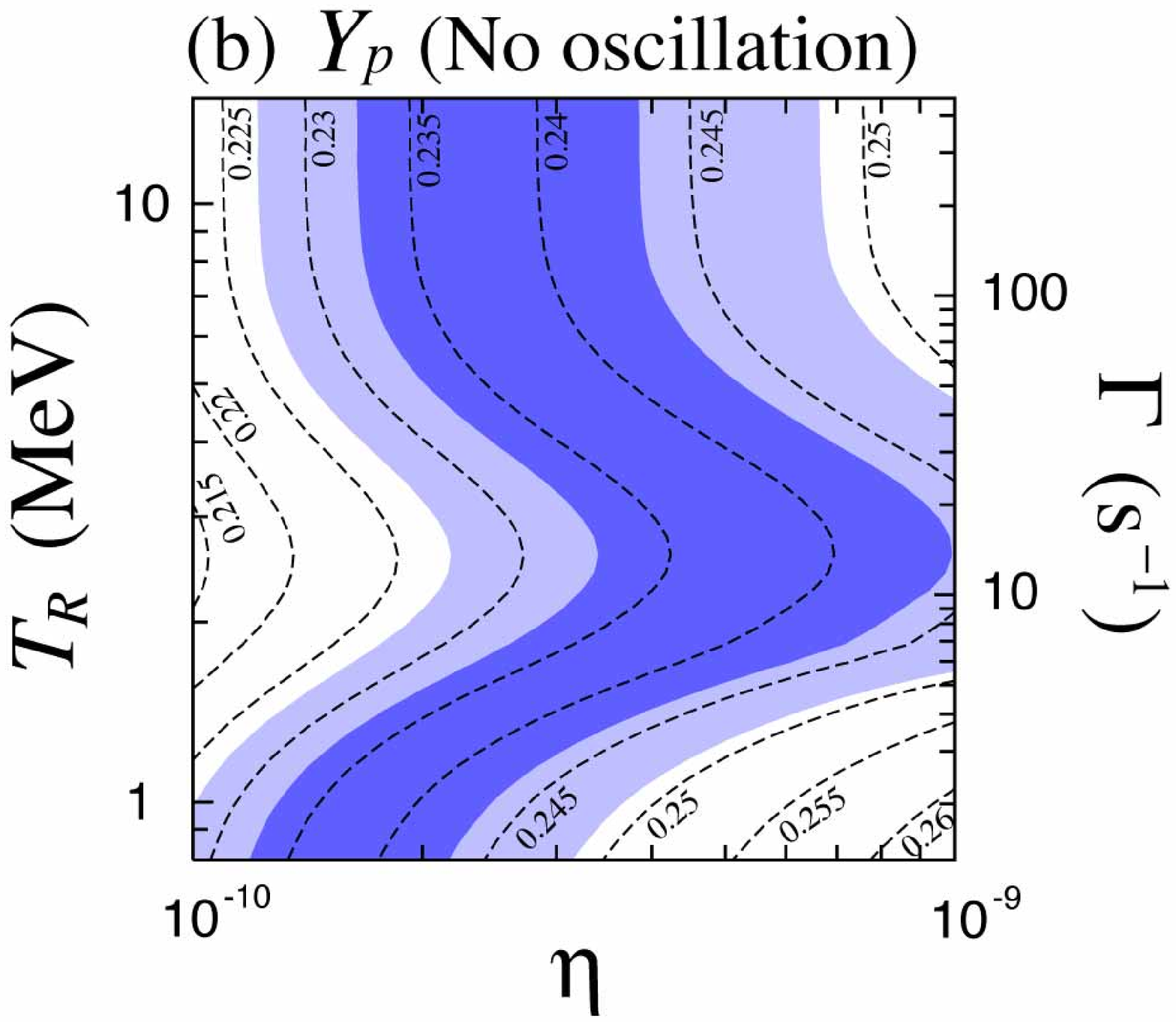} \\
\includegraphics[width=7.5cm]{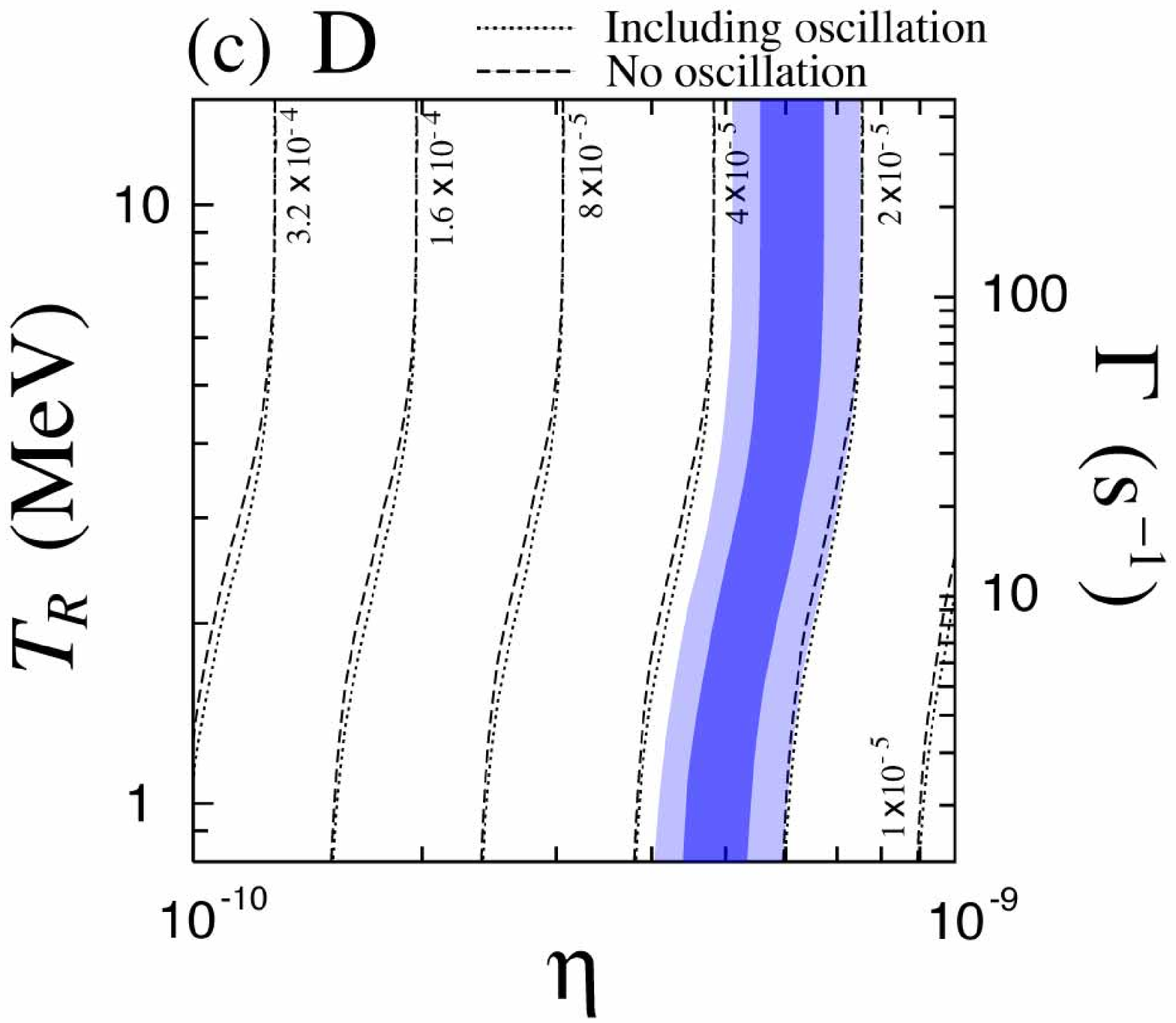} &
\includegraphics[width=7.5cm]{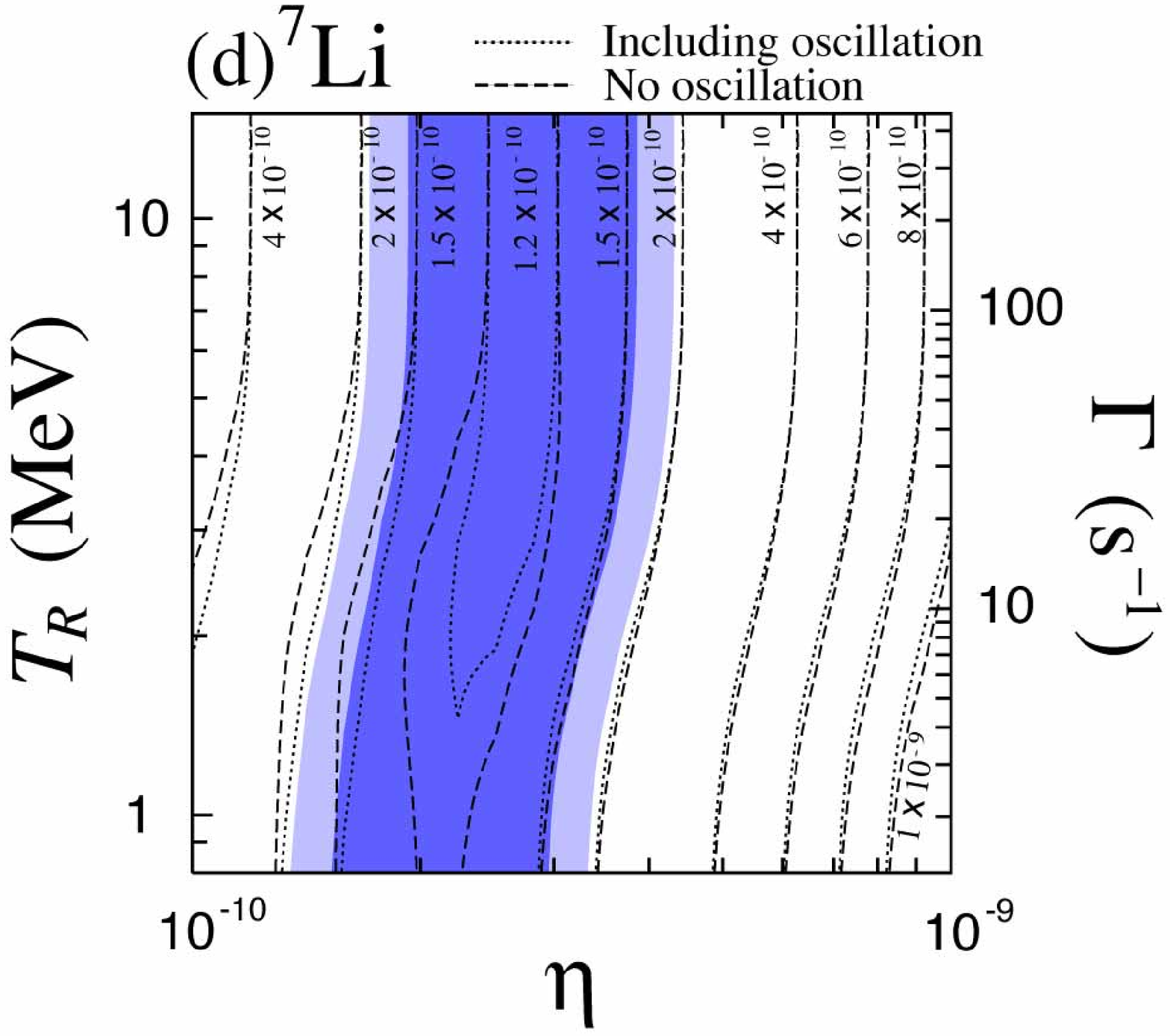} \\
\end{tabular}
\caption{
Contour plots for the light element abundances. $^4$He mass fraction
is plotted in (a) with the oscillations and in (b) without. D and
$^7$Li are plotted respectively in (c) and (d) where dotted lines
express the case with the oscillations and dashed lines express the
case without. Shaded areas represent uncertainties in the observed
abundances expressed in eqs.~(\ref{eq:obs_He_FO}) $\sim$
(\ref{eq:obs_Li_R}) (for D and $^7$Li, they are drawn against the
contours considering the oscillations). Darker areas are for 1
$\sigma$ and lighter for 2 $\sigma$.
}
\label{fig:contour}
\end{center}
\end{figure}

\begin{figure}
\begin{center}
\includegraphics[width=12cm]{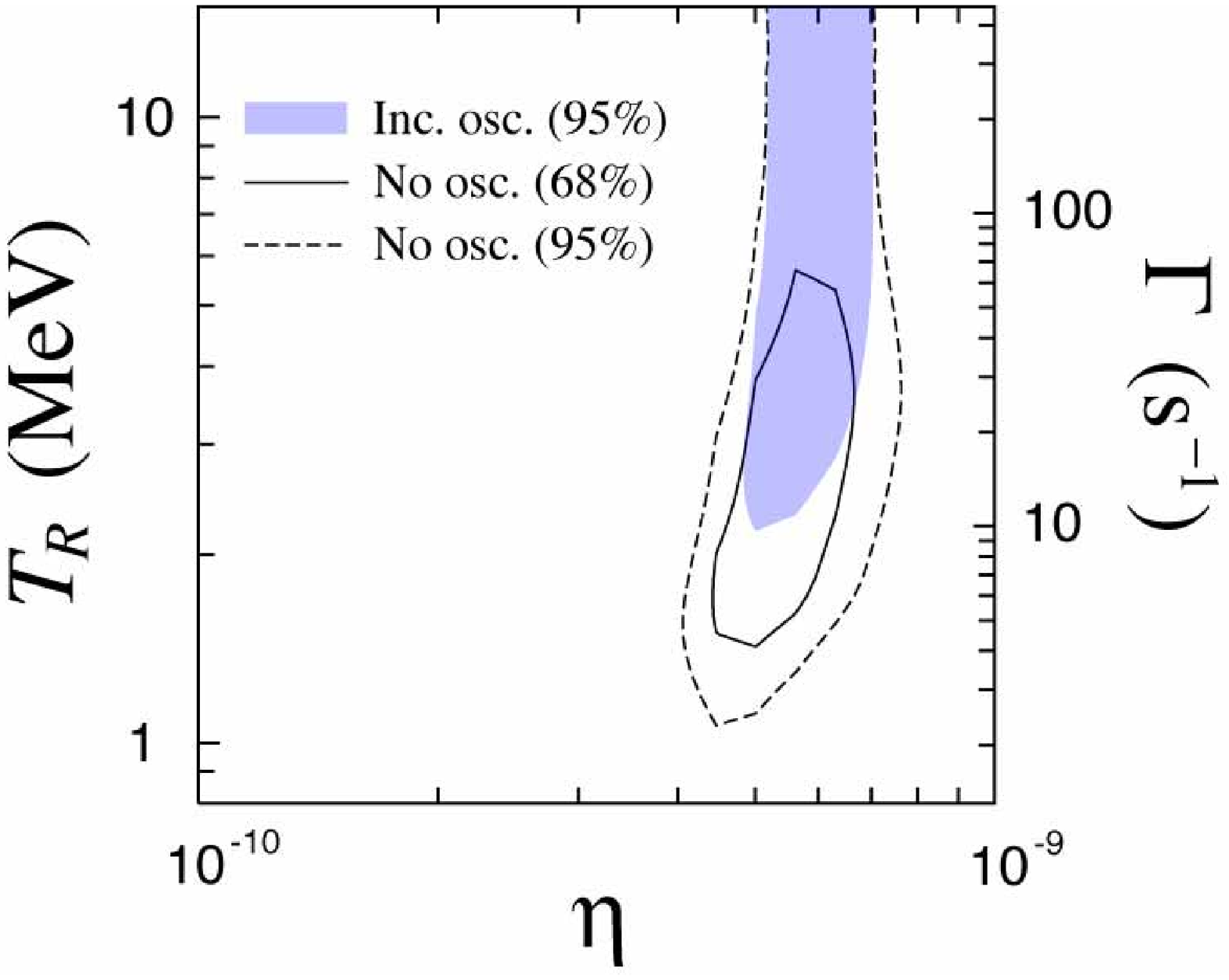} 
\caption{%
$\chi^2$ contour plot using data of D and $^4$He. For no oscillation
case, the allowed regions at 68\% and 95\% confidence levels are drawn
with solid and dashed lines. For the case with oscillations, the 68\%
allowed region does not appear and only the 95\% region is indicated
by the shaded area.
}
\label{fig:chi2contour_bbn}
\end{center}
\end{figure}

On comparing the predictions of low reheating temperature scenario
with the observed abundances, we need to vary the baryon-to-photon
ratio, $\eta$, which is the input parameter for the standard BBN
calculation, in addition to $T_R$. In Fig.~\ref{fig:contour} we show
contour plots for abundances of light elements, D, $^4$He and $^7$Li,
against $\eta$ and $T_R$. Since contours tend to be parallel to each
other, we see that how abundances vary with respect to $T_R$ has
little dependence on $\eta$. In particular, for $^4$He, features found in
Fig.~\ref{fig:TR_He4} seem to appear at every $\eta$. We notice, in
Fig.~\ref{fig:contour} (c) and (d), that the oscillations almost do
not make difference for D and $^7$Li abundances. In the figure, we
also indicate observed values taken from Ref.~\cite{Fields1998} for
$^4$He, from Ref.~\cite{Kirkman:2003uv} for D, and from
Ref.~\cite{Ryan2000} for $^7$Li:
\begin{eqnarray}
\label{he}
Y_p &=& 0.238 \pm 0.002 \pm 0.005, \label{eq:obs_He_FO}\\
\label{d}
{\rm D/H} &=& 2.78^{+0.44}_{-0.38} \times 10^{-5}, \label{eq:obs_D}\\
\label{li}
{\rm ^7Li/H} &=& 1.23^{+0.68}_{-0.32} \times 10^{-10}\quad (95\%) \label{eq:obs_Li_R}
\end{eqnarray}
In Eq.~(\ref{eq:obs_He_FO}), the first uncertainty is statistical and
the second one is systematic. Their root-mean-square, $[{\rm
(stat.)}^2 + {\rm (syst.)}^2]^{1/2}$, is adopted as overall 1$\sigma$
error. In this paper, we do not consider the $^7$Li data since its
systematic error is under debate at present, but show it just for
reference~\footnote{
It is known that the baryon density derived from \EQ{he} is somewhat
lower than one from \EQ{d}. It was widely believed that $N_\nu < 3$
decreases $Y_p$ and ameliorates this tension (see e.g.,
Ref.~\cite{Ichikawa:2004ju}). However we now know this is not feasible
simply by lowering the reheating temperature if we properly take into
account neutrino oscillations.
}.

We immediately realize from Fig.~\ref{fig:contour} (a), (b), (c) that
inclusion of the oscillations leaves smaller room for the low
reheating temperature scenario. In other words, the parameter region
allowed from D and $^4$He measurements is smaller for the case with
the neutrino oscillation. We can see it more clearly by
$\chi^2$-analysis whose results are shown in
Fig.~\ref{fig:chi2contour_bbn}. The lower bound on $T_R$ at 95\%
confidence level in the $\eta$-$T_R$ plane is 1 MeV for the case of no
oscillations but tightened to be 2 MeV for the case incorporating the
oscillations
~\footnote{
Recently, analysis of the $^4$He abundance by Ref.~\cite{Olive:2004kq} suggests $Y_p = 0.249 \pm 0.009$
\cite{Cyburt:2004yc}. This is higher than the value of \EQ{he} mainly due to the different treatments of 
stellar absorption. Although, at present, such large uncertainty does not allow us to derive
any meaningful lower bound on $T_R$. However, higher $Y_p$ is interesting for MeV-scale reheating
scenario. Should future research yield $Y_p > 0.25$, $T_R \sim O({\rm MeV})$ would be favored.
}.

\section{Conclusion} \label{sec:conclusion}

In this paper we have investigated the MeV-scale reheating scenario
wherein the thermalization of neutrinos could be insufficient.  We
have paid particular attention to the oscillation effects on the
thermalization processes of neutrinos, and solved numerically the
momentum dependent Boltzmann equations for neutrino density matrix,
fully taking account of neutrino oscillations.  In contrast to the
widespread picture, we have found that $^4$He abundance does increase
while the effective neutrino number $N_\nu$ decreases. The reason is
simple; the neutrino oscillations reduce the number density of
$\nu_e$, due to which the neutron-proton transformation decouples
earlier. This effect cancels and even overcomes that of the decrease in
the expansion rate; only the latter effect has been usually taken into
account when discussing the effect of $N_\nu$ on the light-element
abundances.  Therefore we would like to stress that it is
indispensable to take into consideration the oscillation effects, to
set a lower bound on the reheating temperature by using the BBN.  As a
reference value, we quote our results; $T_{RH} \gtrsim 2\,{\rm MeV}$
or equivalently $N_\nu \gtrsim 1.2$ obtained by using the
observational data on the $^4$He and D abundances.

What are then the distinct predictions of the MeV-scale reheating?
Clearly, they are: both larger $Y_p$ {\it and} smaller $N_\nu$
compared to their standard values; if both the observed $Y_p$ and
$N_\nu$ suggest the same $T_R$ by the relations shown in
Figs.~\ref{fig:Nnu_compare} and \ref{fig:TR_He4}, they would serve as
decisive evidence for the MeV-scale reheating~\footnote{According to Ref.~\cite{Trotta:2003xg},
we can determine both $Y_p$ and $N_\nu$ with future CMB observations
such as Planck.}.

At last, let us comment on the validity and possible extension of the
present work.  As explained in section \ref{sec:thermalization}, we
have neglected the self-interactions of neutrinos.  Such
simplification is considered to be valid due to the following reason.
Since self-interactions cannot change the total energy stored in
the neutrino sector,  they  affect only the momentum distribution of neutrinos. 
On the other hand, it should be noted that we have taken into consideration 
the neutrino-electron ($\nu e$) scattering, which also shifts the neutrino momentum distribution toward kinetic
equilibrium at the rate of the same order of magnitude as the
$\nu \nu$ scattering. However we have checked that our results
do not change at all even if we increase the $\nu e$ scattering rate a few times larger than the
standard one.
Considering that the $\nu \nu$ scattering rate is further suppressed due to
the deficit in the neutrino number, we are sure that the self-interactions have only
a minor effect in the neutrino momentum distribution. 
Still, the self-interactions
have a potential effect on the number density of $\nu_e$ through e.g.,
$\nu_e
\bar{\nu}_e \leftrightarrow \nu_{\mu(\tau)} \bar{\nu}_{\mu(\tau)}$.
Furthermore, nonzero $\theta_{13}$ can have a similar effect; in this
case it is necessary to perform three generation analysis.
Nevertheless we believe that our main conclusion is robust,
since these extensions, too, are expected to decrease the number
density of $\nu_e$, further increasing the $^4$He abundance. Of course
the quantitative improvement should be necessary and the full analysis
on these points will be presented elsewhere~\cite{in_prep}.

{\it Acknowledgments.}---
  F.T.  would like to thank the Japan Society for
Promotion of Science for financial support.


\end{document}